Spin Dynamics in the Second Subband of a Quasi Two Dimensional System Studied in a Single Barrier Heterostructure by Time Resolved Kerr Rotation

F. Zhang, H. Z. Zheng*, Y. Ji, J. Liu, G. R. Li

The State Key Laboratory for Superlattices and Microstructures, Institute of Semiconductors, Chinese Academy of Sciences, P. O. Box 912, Beijing 100083

By biasing a single barrier heterostructure with a 500nm-thick GaAs layer as the absorption layer, the spin dynamics for both of the first and second subband near the AlAs barrier are examined. We find that when simultaneously scanning the photon energy of both the probe and pump beams, a sign reversal of the Kerr rotation (KR) takes place as long as the probe photons break away the first subband, and start to probe the second subband. This novel feature, while stemming from the exchange interaction, has been used to unambiguously distinguish the different spin dynamics ($T_2^{1*}$ and $T_2^{2*}$) for the first $E_1$ and second $E_2$ subbands under the different conditions by their KR signs (negative for 1st and positive for 2nd). By scanning the wavelength towards the short wavelength in the zero magnetic field, $T_2^{1*}$ decreases in accordance with the D'yakonov-Perel' (DP) spin decoherence mechanism. At 803nm, $T_2^{2*}$ (450ps) becomes ten times longer than $T_2^{1*}$ (50ps). However, the value of $T_2^{2*}$ at 803nm is roughly the same as the value of $T_2^{1*}$ at 815nm. A new feature has been disclosed at the wavelength of 811nm under the bias of -0.3V (807nm under the bias of -0.6V) that the spin coherence times ($T_2^{1*}$ and $T_2^{2*}$) and the effective $g^*$ factors ($|g^*(E_1)|$ and $|g^*(E_2)|$) all display a sudden change, due to the "resonant" spin exchange coupling between two spin opposite bands.

* To whom the correspondence should be addressed. E-mail: hzzheng@red.semi.ac.cn

The behaviors of spin coherence in bulk semiconductors and their low-dimensional quantum structures have been extensively studied [1] in order to make it feasible that the spin degree of freedom can be employed as an alternative carrier of information in the next generation of electronics. Most experimental investigations have focused on the dynamics of spin decoherence in ground states [2]. However, few have studied the spin coherence in excited states. It was theoretically predicted that due to strong inter-subband scattering, the spin decoherence rate of electrons in ground and excited subbands were almost identical, despite the large difference in the

D'yakonov-Perel' (DP) terms of different subbands[3]. Meanwhile, the spin decoherence rate in the second subband was found to be much slower than in the first subband due to a small spin–orbit splitting at small Fermi wave vector in the weakly occupied second subband, as revealed by the measurement of a positive magnetoresistance at low magnetic fields, which was induced by spin-orbital scattering [4]. To clarify the spin dynamics in the second subband and the possible interplay between the 1st and 2nd subbands, it is desirable to directly observe the temporal spin evolutions in the second subband as in the first subband.

In this work, the population in the second subband is created by drifting the spin-polarized electrons, excited in the 500nm-thick GaAs layer by a circularly polarized pump pulse, into the vicinity of an AlAs barrier in a single barrier heterostructure, thus forming a quasi two dimentional electron system (Q2DES). Because the renormalized single particle energy due to the exchange interaction depends on the population, it thus makes the majority and minority spin bands in both the first and second subbands descend differently. As a result, the majority spin band of $E_2$ may overlap with the minority spin band of $E_1$ in certain "resonant" energy regime. In this manner, we can unambiguously distinguish the different spin dynamics for $E_1$ and $E_2$ by their KR signs (negative for $E_1$ and positive for $E_2$), and studied their dynamics under the different conditions.

The sample structure for this investigation was a single-barrier tunneling diode, grown by molecular beam epitaxy (MBE) with a thick intrinsic GaAs as the absorption layer. The layer structures, sample preparation and measurement configuration for the time resolved Kerr rotation (TRKR) were the same as in Ref. [5]

To examine the dynamics of the spin decoherence in the second subband, the spin-polarized electrons, excited in the 500nm-thick GaAs layer by a circularly polarized pump pulse, should first drift into the vicinity of the AlAs barrier by biasing the structure so that the second subband can be simultaneously populated to some extent with the ground subband. For a certain photon energy, the spin dynamics in the first and second subbands can be detectable simultaneously in our structure, as long as the photon energy of the probe beam fit the transition energy for the interband absorptions from the valence band to both the first and second subbands in the conduction band, which take place in the different spatial regions, as schematically shown in Fig. 1 by the hatched areas underneath the respective bands. In our recent work [5], we demonstrated that, instead of the Rashba and Dresselhaus types, a dynamic spin splitting along the growth direction can be induced in heterostructures, when a population imbalance between two electron spin bands is created by circularly polarized excitation. This is because the single particle energy of an electron will be renormalized due to its exchange interaction with other electrons in interaction electron gas, and the difference of the renormalized energy between the majority and minority spin bands gives rise to an observable dynamic spin splitting. As long as such exchange-interaction-induced spin splitting is large enough to lift the quasi Fermi level ($E_F^-$) in the minority spin band above that ($E_F^+$) in the majority spin band (as depicted by the inset (a) to Fig.1), both the sign of KR and the phase of Larmor precession can be switched or reversed by scanning the wavelengths of both the pump and probe beams simultaneously. Fig.2 gives the KR in an extended wavelength range, which are measured at a fixed probe delay time of 100ps under the biases of 0V, -0.3V, and -0.6V by scanning the wavelength of both the pump and probe beams. After the photon energy becomes larger than the fundamental band gap (at ~816.6nm for -0.3V, -0.6V, and ~816.7nm for 0V), the sign reversal of the KR from the positive to negative takes place at the wavelength shorter than 816.1nm and 815.5nm for the biases of -0.3V and -0.6V, respectively. As verified in Ref. [5],

$$\theta_K \propto \left\{ \sqrt{\hbar\omega - E_g^+} \left( 1 + e^{\frac{\hbar\omega - E_g^+ - \mu_F^+}{kT}} \right)^{-1} - \sqrt{\hbar\omega - E_g^-} \left( 1 + e^{\frac{\hbar\omega - E_g^- - \mu_F^-}{kT}} \right)^{-1} \right\} \quad (1).$$

Here, $E_g^+$ and $E_g^-$ are the effective band gaps for the split majority and minority spin bands with $E_g^+$ smaller than $E_g^-$. $\mu_F^+$ and $\mu_F^-$ are the respective quasi chemical potentials, measured from the respective bottoms of the majority and minority spin bands ($\mu_F^+ > \mu_F^-$). Then, the sign reversal of the KR occurs when the population difference between the majority and minority spin subbands (e.g., in the first subband $E_1$), detected by the probe photons, is inversed by scanning the photon energy $\hbar\omega$ across the quasi Fermi level $E_{F1}^+$ ($E_{F1}^+ = E_{g1}^+ + \mu_{F1}^+$ and $E_{F1}^- = E_{g1}^- + \mu_{F1}^-$) in the majority spin band (where spin down states reside in the present case). A similar spin splitting happens to the second subband $E_2$ as well with $E_{F2}^-$ either below or above $E_{F2}^+$, as shown by the inset (b) to Fig. 1. From this physical understanding, one expects that by scanning the photon energy $\hbar\omega$ further upwards, another population inversion between the minority spin band in $E_1$ and the majority spin band in $E_2$ should occur when the probe photons break away from the first subband and start to probe the second subband, as illustrated in Fig. 1. Next, we use this feature to trace the appearance of the KR from the second subband. As seen from Fig. 2, the KR for the biases of -0.3V and -0.6V switch their sign once more from negative back to positive at the wavelengths of about 810nm and 806nm, respectively.

In Fig. 3, the time evolutions of the KR are measured at wavelengths of 815nm, 811nm, 809nm, and 807nm under both zero (the left panel) and 2T (the right panel) field for the case of -0.3V. When scanning the wavelength of the probe beam from 815nm to 807 nm, a sign reversal process at the zero field (or a phase reversal process accompanied by quantum beatings at 2T) takes place around the wavelength of 810nm. This fact provides clear evidence that there must be two KR transient processes with the opposite signs involved in our TRKR measurements. To understand its physical origin, on one hand, the renormalized single particle energy due to the exchange interaction is a negative correction term, and its magnitude depends on the population. That makes the majority and minority spin bands in both the first and second subbands descend differently. As a result, the majority spin band of $E_2$ may overlap with the minority spin band of $E_1$ in certain "resonant" energy ranges, as depicted by the horizontal line in Fig. 1 (which is the energy baseline with respect to the photo-exxcitation). On the other hand, the effective transition regions in the space for $E_1$ and $E_2$ subbands, as indicated by the hatched areas underneath both of $E_1$ and $E_2$, may also possibly share the same transition energy (labeled by two arrows in Fig.1). As a result, quantum beating occurs naturally between two Larmor oscillations from the first and second subbands with slightly different effective masses.

As the wavelength scans from 816nm to 803nm, the photon energy falls in the gap between $\mu_F^+$ and $\mu_F^-$ in the $E_1$ subband. Judging from Eq. (1), one knows that the KR from $E_1$ has a negative sign. Meanwhile the KR from $E_2$ is positive and increases in its magnitude when the photon energy is higher than $E_{g2}^+$. This superposition is perfectly in accordance with our observations in Fig. 2, leading a sign (or phase) reversal at about 809nm~810nm under -0.3V bias.. Similar behaviors have also been seen at -0.6V, where a sign (or a phase) reversal process at the zero field takes place around a wavelength of 806nm. In order to properly extract two different spin coherence times and the effective $g^*$ factors from the data, we have used two different temporal exponentials for the first (i=1) and second (i=2) subbands, which are in the forms of $C^i \exp\left[-(t-t_0)/T_2^{i*}\right]$ at the zero magnetic field and $C^i \exp\left[-(t-t_0)/T_2^{i*}\right]\cos\left[\omega_L^i(t-t_0)\right]$ at 2T, to fit the measured data. The fitting temporal evolutions for the zero and 2T fields are all in very good agreement with the measured data and are hardly discernible from each other, as shown in Fig. 3. Fig. 4 gives the wavelength dependences of the spin coherent times for $E_1$ and $E_2$ under various conditions. Since the KR sign, or phase, in the wavelength range studied here is confirmed to be negative for the first subband and positive for the second subband, one can unambiguously distinguish the spin dynamics for the $E_1$ and $E_2$ bands by the their KR sign.

Let us first look at the spin coherent times $T_2^*$ for the case of the -0.3V and zero magnetic field, as shown in Fig. 4a. With scanning the wavelength of the probe beam towards the short wavelength side, as an overall trend, the spin coherent time $T_2^{1*}$ of $E_1$ decreases from 450ps at 815nm to 50ps at 803nm. But it displays a sudden drop at the wavelength of 811nm. The former can be explained in the framework of the D'yakonov-Perel' (DP) decoherence mechanism [3], which increases with the wave vector in the plane, or equivalently with the photon energy in the present case. The latter discloses a new feature. At the resonant wavelength of 811nm, where the photon energy simultaneously probes the dynamics near $E_{F1}^-$ of the spin minority band in $E_1$ and that close to the bottom of the spin majority band in $E_2$, the spins near the $E_{F1}^-$ in $E_1$ will suffer additional exchange scattering from the spin majority band in $E_2$. Such scattering only plays an important role near the Fermi level $E_{F1}^-$, leading to the observed sudden drop in $T_2^{1*}$ at 810nm. For the spin coherence time $T_2^{2*}$ in $E_2$, it remains very low at wavelengths longer than 810nm, and then dramatically enhances afterwards. This feature is not well understood yet. As seen from Fig. 1, the photons with their wavelengths longer than 810nm are only able to detect the spins in $E_2$, which spatially dwell in the vicinity of the left boundary of Q2DES. These spins have a relatively lower carrier density. We hypothesize that they may suffer the exchange scattering rather effectively from the spin minority band of $E_1$. When the wavelength becomes shorter than 810nm, the spins with higher carrier density in the central part of the triangle-like quantum well make the main contribution to the KR. The inter-subband spin exchange scattering may be suppressed due to the decrease in the number of available empty final states. That, together with still small DP effect in $E_2$, gives rise to a longer spin coherence time $T_2^{2*}$ at the wavelengths shorter than 810nm.

Both $T_2^{1*}$ and $T_2^{2*}$ were measured under the bias of -0.3V and the field of 2T in Fig. 4b. They show a similar variation, with sudden changes appearing at 810nm for both $T_2^{1*}$ and $T_2^{2*}$. However, because of inhomogeneous decoherence events stemming from either the fluctuation in the local effective $g^*$ factor or the local magnetic field (e.g. the Rashba and Dresselhaus fields), the differences between $T_2^{1*}$ and $T_2^{2*}$ are greatly suppressed, especially on the short wavelength side. For the case of -0.6V and 2T in Fig. 4c, the minimum of both $T_2^{1*}$ and $T_2^{2*}$ shifts to 807nm, where the phase reversal of the Larmor precessions occurs. Compared with the case of -0.3V and 2T, the increased splitting between $E_2$ and $E_1$ by biasing seems to accelerate the intersubband scattering, and tends to equalize $T_2^{2*}$ with $T_2^{1*}$ more efficiently.

The effective $g^*$ factors for $E_1$ and $E_2$ are also extracted in Fig. 4d and 4e for the biases of -0.3V and -0.6V, respectively. Fig.3d shows that $\left|g^*(E_2)\right|$ decreases from 0.4575 at 815nm to 0.440 at 805nm, with a steep drop at about 811nm. Meanwhile, $\left|g^*(E_1)\right|$ remains at 0.440 when the wavelength is longer than 811nm, then jumps to 0.445 in the range from 809nm to 803nm. Following $\vec{k}\cdot\vec{p}$ perturbation theory with the spin-orbit interaction included, the Lande factor near the band edge of GaAs has the form of $-0.44+6.3E$ for the 3D case, and $-0.377+4.5E$ for the 2D case (where E is in the unit of eV) [6]. As a result, the Lande factor will generally show a decreasing trend as the photon energy of the probe beam increases. The variation of $|g^*(E_2)|$ satisfies the mentioned trend. However, both the steep falling of $|g^*(E_2)|$ and the sudden jumping of $|g^*(E_1)|$ from 0.440 to 0.445 are still difficult to understand. The latter may still be related to the fact that the spins detected by probe photons are spatially shifted from the left boundary to the central part of the triangle-like quantum well. It also seems that at the wavelength of about 811nm, the resonant exchange spin coupling between the spin minority band of $E_1$ and the spin majority band of $E_2$, as mentioned previously, tends to cause band mixing to some extent. This needs to be clarified by further work.

In conclusion, by biasing a single barrier heterostructure with a 500nm-thick GaAs layer as the absorption layer, the spin-polarized electrons, excited in the 500nm-thick GaAs layer by circularly polarized pump pulse, are drifted into the vicinity of AlAs barrier so that the second subband can be simultaneously populated to some extent with the ground subband. By simultaneously scanning the photon energy of both the probe and pump beams, the sign reversal of the Kerr rotation takes place as long as the probe photons break away the first subband, and start to probe the second subband. This novel feature has been used to unambiguously distinguish and study the different spin dynamics ($T_2^{1*}$ and $T_2^{2*}$) for the first and second subbands under the different conditions. In the zero magnetic field, by scanning the wavelength towards the short wavelength side, $T_2^{1*}$ decreases as the wave vector is gradually enlarged in accordance with the DP spin decoherence mechanism. Eventually, $T_2^{2*}$ becomes ten times longer than $T_2^{1*}$, as probed at 803nm, indicating that the DP term in $E_2$ is much less effective than in $E_1$ due to the smaller wave vector there. However, the value of $T_2^{2*}$ at 803nm is roughly the same as the value of $T_2^{1*}$ at 815nm. Under the magnetic field of 2T, as the inhomogeneous decoherence events stemming from either the fluctuation in the local effective $g^*$ factor or the local magnetic field set in (e.g. the Rashba and Dresselhaus fields), $T_2^{1*}$ and $T_2^{2*}$ tend to be equalized to a low value of 200ps (150ps) on the short wavelength side for -0.3V (-0.6V). A new feature has been disclosed at the wavelength of 811nm under the bias of -0.3V (807nm under the bias of -0.6V) that the spin coherence times ($T_2^{1*}$ and $T_2^{2*}$) and the effective $g^*$ factors ($|g^*(E_1)|$ and $|g^*(E_2)|$) all display a sudden change. That is attributed to the "resonant" spin exchange coupling between two spin opposite bands, occurring when the probe photons simultaneously detect the minority spin in $E_1$ and the majority spin in $E_2$. Our result give a complete picture of the spin dynamics in the second subband of

Q2DES.

The authors would like to thank Z. C. Niu for sample growth. This work was in part supported by National Basic Research Program of China No 2006CB932801and No2007CB924904, and also by Special Research Programs of Chinese Academy of Sciences.

Figure Captions

Fig.1 energy band profile near an AlAs barrier with two subbands indicated. The hatched areas under the two subbands indicate the space where interband transitions may occur. Two arrows inside two hatched areas represent the interband absorptions to $E_2$ and $E_1$, excited by same photon energy. Horizontal line, strentching to both inset (a) and (b), is the energy baseline set by photon energy. Inset (a) and (b) sketch two renormalized spin opposite bands for two subbands.

Fig.2 KR are measured by simultaneously scanning both pump and probe pulses for 0V, -0.3V, and -0.6V

Fig. 3 KR temporal evolutions under both zero (left panel) and 2T (right panel) fields measured at different wavelengths of 815nm, 811nm, 809nm, and 807nm

Fig. 4 (a), (b) and (c) Spin coherence times for 1[st] and 2[nd] subbands, as the wavelength varies for three different conditions: -0.3V 0T, -0.3V 2T, and -0.6V 2T. The points are experimental results, and the lines are best fittings to the points. (d) and (e) Effective $g^*$ factors for 1[st] and 2[nd] subbands, extracted from data in Fig. 2, plotted as a function of wavelength

Fig. 1

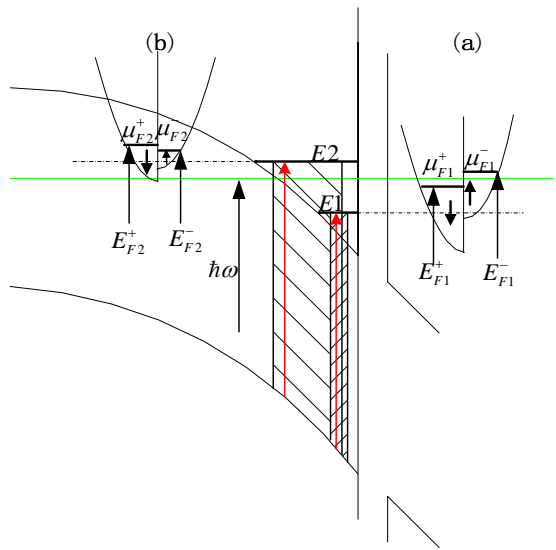

Fig. 2

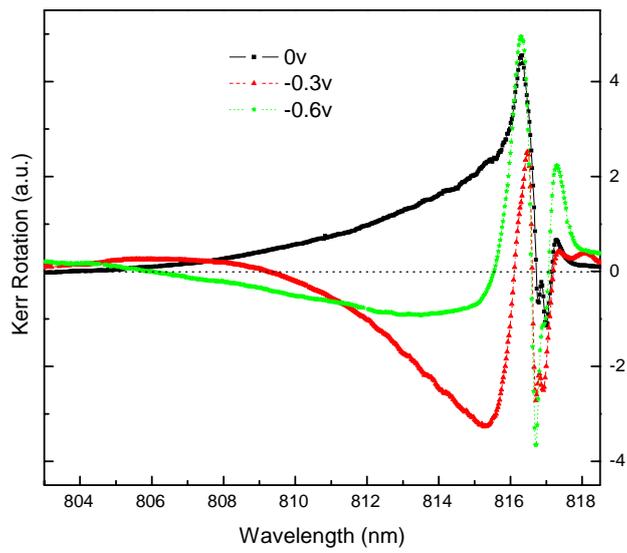

Fig. 3

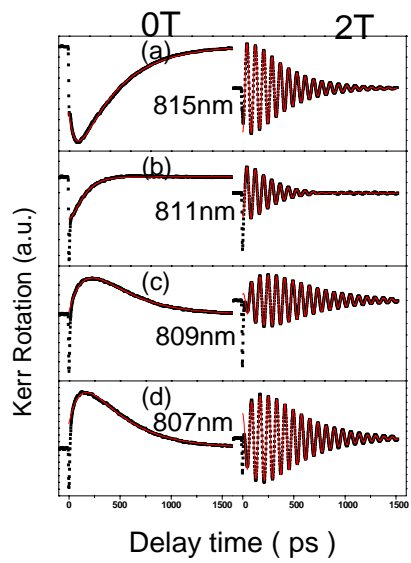

Fig. 4

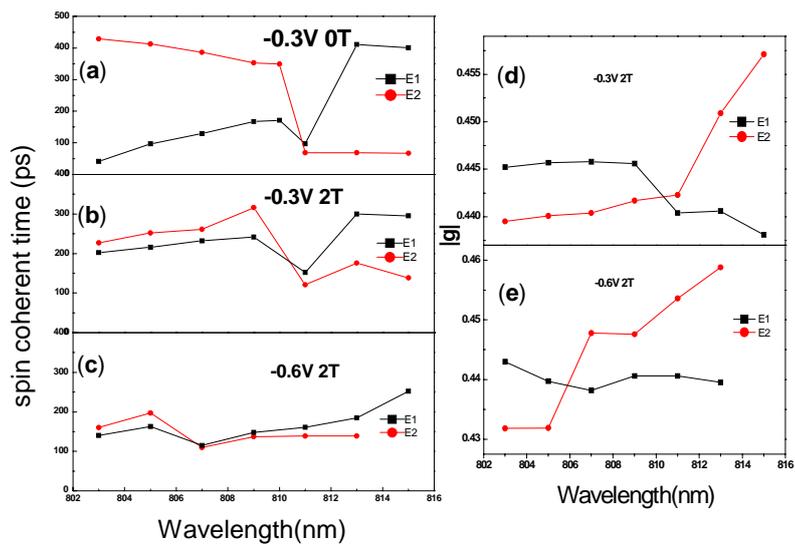